\documentstyle[prl,aps,epsfig,preprint,floats]{revtex}
\begin{document}
\normalsize \draft \title{Weakly correlated electrons on a square lattice: a
renormalization group theory}

\author{D. Zanchi}
\address{Laboratoire de Physique des Solides, Universit\'e Paris-Sud,
91405 Orsay, France, and Institut f\"ur Theoretische Physik der Freien
Universit\"at Berlin, Arnimallee 14, 14195 Berlin, Germany}

\author{H. J. Schulz}
\address{Laboratoire de Physique des Solides, Universit\'e Paris-Sud,
91405 Orsay, France}
%\date{\today }
\maketitle

\widetext

\begin{abstract}
We study the weakly interacting Hubbard model on the square lattice using a
one--loop renormalization group approach. The transition temperature $T_c$
between the metallic and (nearly) ordered states is found. In the parquet
regime, $(T_c\gg |\mu|)$, the dominant correlations at temperatures below
$T_c$ are antiferromagnetic while in the BCS regime $(T_c\ll |\mu|)$ at
$T_c$ the d-wave singlet pairing susceptibility is most divergent.
\end{abstract}


%\begin{multicols}{2}

%\narrowtext

%\newpage

A theoretical understanding of systems with two different kinds of
coexisting and possibly competing instabilities of density--wave (DW) and
superconducting (SC) type remains one of the central problems of the theory
of interacting fermions. In particular, the question of a purely electronic
mechanism of superconductivity, induced by an incipient instability of the
density--wave type, remains difficult if one wants to go beyond the most
qualitative level. Typically the existence and strength of the DW
correlations and their coupling to superconducting pairing depend strongly
on the dimensionality and on the geometry of the Fermi Surface.

Experimentally, superconductivity in the vicinity of an insulating and/or
spin density wave state is a well--known property of the phase diagrams for
several families of fermion systems: for example, in the
quasi--one--dimensional Bechgaard salts,\cite{JerS} superconductivity
(possibly of d--type \cite{Takigawa}) replaces an SDW state as one increases
pressure, due to two effects: decreasing of the nesting properties of the
Fermi surface, and suppression of umklapp scattering between electrons. The
phase diagram of the quasi--two--dimensional organic superconductors of the
ET family also shows a pressure induced SDW--SC transition
\cite{Hadrienetal}, but the symmetry of the SC order parameter is not yet
clear.\cite{mu-sr} In the high--$T_c$ superconductors a few percent doping
transforms an insulating antiferromagnet into a superconductor\cite{books},
probably of the $d_{x^2-y^2}$ symmetry.\cite{d-wave} In two dimensions the
simplest model showing an insulating antiferromagnetic state already at weak
coupling is the Hubbard model, often considered as a ``minimal model'' for
high-$T_c$ superconductors.\cite{Anderson_87} Although extensively studied,
the question of doping induced superconductivity in the vicinity of the
antiferromagnetic state in this model still remains unanswered.  In the weak
coupling limit the problem of the interdependence of different kinds of
instabilities can be treated using renormalization group methods
\cite{Shankar,ZS_prb}, as has been successfully done for the case of
quasi--one--dimensional systems\cite{Rev1,Bourbonnais}.

In the present paper we present a one--loop renormalization group analysis
of the Hubbard model. Because of its perturbative nature this gives a
quantitatively correct phase diagram only for weak coupling and does not
allow us to make precise predictions about realistic systems where the
coupling is rather strong. However, we are able to describe the physical
mechanisms which lead to different instabilities of the model.  Provided
that there is no transition to a qualitatively different regime for strong
repulsion, this should give at least some qualitative insight even outside
the strictly perturbative regime, at least for the effective low energy
model \cite{Shankar}. Recently we have formulated one--loop
renormalization--group equations for the Hubbard model,\cite{ZS_prb} based
on the requirement that the vertex (the effective interaction) must be
invariant under changes of the energy cutoff $\Lambda$ about the Fermi
surface. This way of renormalization is known as field theory
approach.\cite{Shankar} We distinguished two regimes, separated by a
crossover energy $T_{co}$ which is a function of the chemical potential
$\mu$.  In the parquet regime, $\Lambda > T_{co}$, the contributions to the
renormalization from the particle-particle and particle-hole loops are both
important. In the BCS regime, $\Lambda < T_{co}$, only the particle-particle
loop contributes to the flow, while the particle-hole part is negligibly
small. In the parquet regime both loops behave like $\log ^2 \Lambda$, while
in the BCS regime, the p-p loop crosses over to $\log \Lambda$ and the p-h
loop disappears with some positive power of $\Lambda$.  For the exactly
half--filled case, the parquet equations gave an instability in the
antiferromagnetic channel \cite{DzYak}. The same result was found within a
simple scaling theory \cite{HJS} which takes into account only processes
between electrons at the van Hove points in the corners of the square Fermi
surface. Introducing the chemical potential as a cutoff for the p-h term of
the flow equation this theory also gives a transition to d-wave
superconductivity.

\begin{figure}
\centerline{\epsfig{file=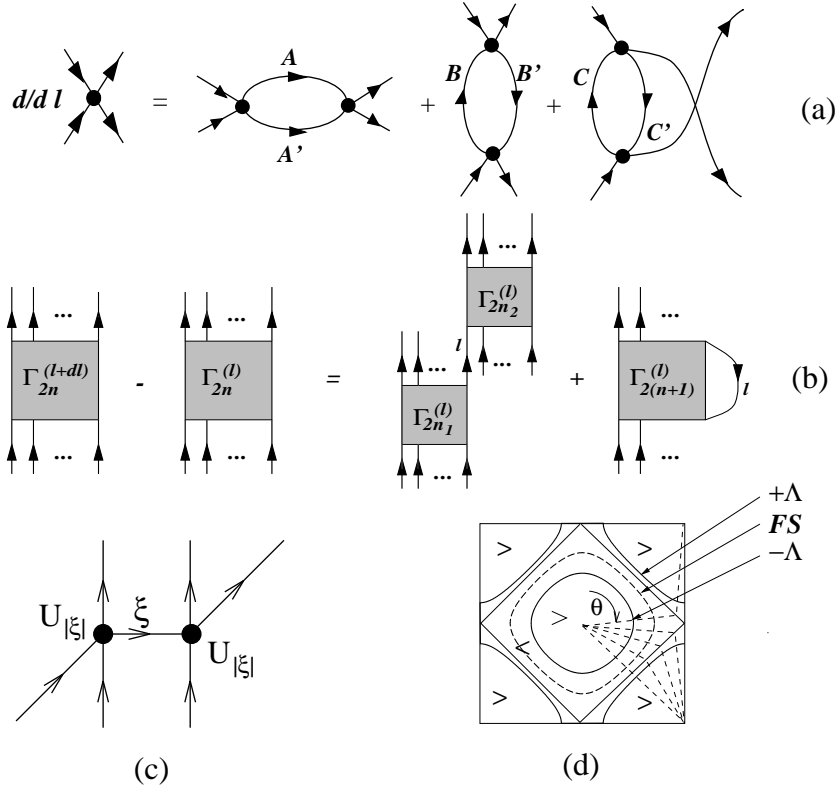,width=11cm}}
\caption{ \narrowtext (a) The one-loop renormalization of the vertex. (b)
The Polchinski equation describes the exact flow of all vertices.  The
contractions labeled with $l$ are integrated over a phase space shell of
energy width $\Lambda d \Lambda$.  (c) The 6--leg diagram generated by the
first term in Polchinski's equation.  The second term applied to it
generates the loops for the renormalization of the 4--leg vertex.  (d) The
organization of the mode elimination for a non--half--filled case.  Dashed
lines are the lines of constant ``angle'' $\theta$.}
\label{Figure 1}
\end{figure}

An important aspect of the renormalization group is that in many cases
$\Lambda$ can be interpreted as the temperature to logarithmic precision.
This approximation consists in renormalizing the vertex at some temperature
$T$ only by virtual processes involving ``quantum'' electrons, those with
energy larger than $T$. Then the vertex depends only on energy variables
greater than $T$ and its dependence on energies inside the area $\pm
\Lambda$ about the Fermi energy is irrelevant.  In that sense the field
theoretic approach can give correct thermodynamics only for
logarithmic--scale--invariant problems, because in that case the internal
fermionic propagators A, A', B, B', C, and C' in fig. \ref{Figure 1}(a) are
all {\it exactly} at the energy $\pm \Lambda$ and, consequently, the vertex
at the step $\Lambda$ of renormalization contains only contributions from
the degrees of freedom {\it outside} the ring $\pm \Lambda$. If there is no
logarithmic scale invariance, as for the case when $\Lambda >T_{co}\neq 0$,
in the field theory approach the lines A,B and C are on the shell, but A',
B' and C' can even lie {\it inside} the ring $\pm \Lambda$, and $\Lambda$
can thus not be considered as an effective temperature. In order to obtain
thermodynamics from the renormalization group we thus have to modify the
bookkeeping of mode elimination in a way so that the modes inside the
temperature ring $\pm T$ about the Fermi energy never enter the integration.
This can be done using the Kadanoff-Wilson mode elimination technique
\cite{Ma} as given by Polchinski's
equations.\cite{Polchinski84,Morris94,these} In the BCS regime, the field
theory approach is valid because there only the p-p channel survives and
scales just as $\log \Lambda$.

In this letter we solve Polchinski's flow equation at the one--loop level,
as shown schematically in fig.\ref{Figure 1}. This allows us to find the
renormalization of the interaction in the nontrivial case when the proximity
of half--filling, via nesting and van Hove singularities, makes both DW and
SC tendencies strong and the critical temperature $T_c$ can be already in
the parquet regime.  The renormalization of different correlation functions
gives us the phase diagram.

In the Kadanoff--Wilson scheme, the one-loop renormalization flow of the
interaction $U(K_1,K_2,K_3,K_4( K_1,K_2,K_3))$ as a function of the three
energy-momenta $K_i\equiv (\omega _i, {\bf k}_i)$ is defined graphically by
fig.\ref{Figure 1}(a), where the propagators A, B, and C are on the shell
$\pm \Lambda$, but A', B', and C' are constrained to run only over states
with $|\xi |\geq \Lambda$.  The interactions in the loops are also to be
taken at the same cutoff as the propagators A', B' and C' and not at the
actual cutoff $\Lambda $. This means that the renormalization of the
interaction is nonlocal in $\Lambda$: this is the price we have to pay to
get thermodynamics from the renormalization group. For the Hubbard model the
initial condition is $U_{\Lambda = \Lambda _0}=U$, where $\Lambda _0$ is the
initial cutoff which we take to be equal to the bandwidth ($=8t$) in order
to take degrees of freedom from the whole Brillouin zone into account. It is
convenient to introduce the logarithmic scale $l\equiv \ln 8t/\Lambda$.

The fact that $U_{\Lambda}$ is a function of six variables makes the
renormalization equations very difficult. Power counting arguments can
provide some drastic simplifications, in a way that allows us to eliminate
from $U_{\Lambda}$ all irrelevant variables, namely the frequencies $\omega
_i$ and energy variables $\xi _{\bf k _i}$ of the interacting particles. A
problem with scaling in the parquet regime with $\mu \neq 0$ is that if one
chooses processes exactly at the Fermi surface as marginally relevant, the
nesting and umklapp scattering processes appear formally as irrelevant,
whereas this irrelevancy only really occurs when $\Lambda \ll |\mu|$, i.e.
in the BCS regime. On the other hand, the irrelevance of the energy
variables allows us to choose as marginally relevant a set of processes
between electrons at any distance from the Fermi surface in the ring $\pm
\Lambda$, not necessarily at the Fermi surface. Thus, in the parquet regime
we keep nesting and umklapp processes even for $\mu \neq 0$ by considering
processes between electrons exactly at the square Fermi surface of the
half--filled case. Up to irrelevant corrections $U$ then depends only on
projections of the momenta on this square, as shown in Fig. 1(d),
i.e. $U_{\Lambda}=U_{\Lambda}(\theta _1,\theta _2,\theta _3)$.  On the other
side of the crossover, in the BCS regime there remains only the BCS
amplitude $V_{\Lambda}(\theta _1,\theta _3)=U_{\Lambda} (\theta _1=\theta
_2+\pi,\theta _3=\theta _4+\pi)$, but now the particles are at the Fermi
surface, and not on the square, since the square is outside the ring $\pm
\Lambda$.

\begin{figure}
\centerline{\epsfig{file=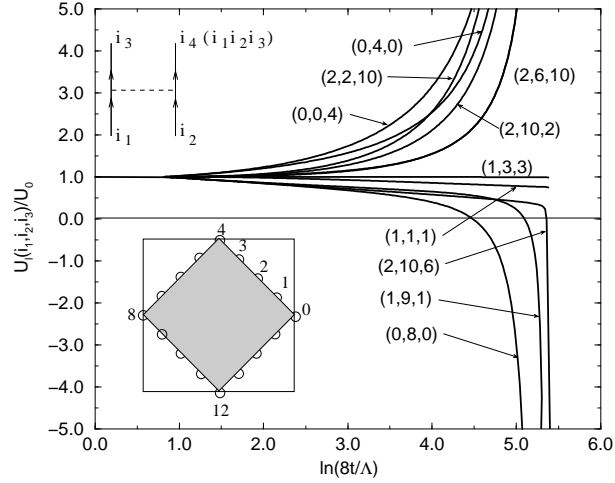,width=9cm}}
\caption{
\narrowtext 
The flow of a few typical (among 93) scattering amplitudes 
for a Fermi surface covered by 16 patches, for chemical potential 
$\mu=8t\exp (-7.8)$ and initial interaction $U=4t/3$.}
\label{Figure 2}
\end{figure}

\begin{figure}
\centerline{\epsfig{file=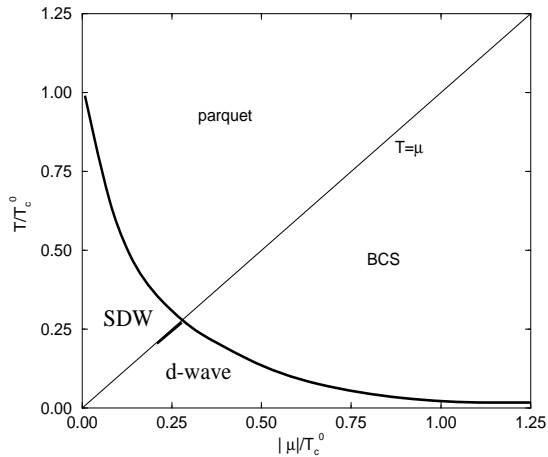,width=8cm,height=7cm}}
\caption{
\narrowtext 
The phase diagram.}
\label{Figure 5}
\end{figure}

The renormalization equation for the interaction as function of $\theta _i$
only is of the same form as in ref.\cite{ZS_prb}, but now with
\begin{equation} \label{Betaee}
\beta _{ee}\{ U,U\}(\theta _1,\theta _2,\theta _3)= 
\Xi \{ U,U\}(\theta _1,\theta _2,\theta _3) +\Xi \{ XU,XU\} 
(\theta _1,\theta _2,\theta _3)
\end{equation}
and
\begin{equation} \label{Betaeh}
\beta _{eh}\{ U_1,U_2\}(\theta _1,\theta _2,\theta _3)=
\Pi \{ U_1,U_2\}(\theta _1,\theta _2,\theta _3) +
\Pi \{ U_1,U_2\}(\theta _3,\theta _4,\theta _1)
,
\end{equation}
where
$$
\Xi \{ U,U\}(\theta _1,\theta _2,\theta _3)
 =\frac{-2}{(2\pi)^2}\sum _{\nu =+,-}
\int d\theta {\cal J}(\nu\Lambda,\theta)
$$
$$
\frac{
\Theta
\left(\nu\xi _{{\bf k}_{\nu}-{\bf q}_{ee}}
 \right) 
\Theta
\left(|\xi _{{\bf k}_{\nu}-{\bf q}_{ee}}|-\Lambda
 \right) 
}{1+\frac{\nu}{\Lambda}\xi _{{\bf k}_{\nu}-{\bf q}_{ee}}}
\times 
$$
\begin{equation} \label{Xi-theta}
\times  U_{|\xi _{{\bf k}_{\nu}-{\bf q}_{ee}}|}(\theta_1,\theta_2,\theta)
U_{|\xi _{{\bf k}_{\nu}-{\bf q}_{ee}}|}(\theta_3,\theta_4,\theta)\; ,
\end{equation}
$$
\Pi \{ U_1,U_2\}(\theta _1,\theta _2,\theta _3)
 =\frac{2}{(2\pi)^2}\sum _{\nu =+,-}
\int d\theta {\cal J}({\nu}\Lambda,\theta)
$$
$$\frac{\Theta
\left(-\nu\xi _{{\bf k}_{\nu}+{\bf q}_{eh}}
 \right) 
\Theta
\left(|\xi _{{\bf k}_{\nu}+{\bf q}_{eh}}|-\Lambda
 \right) }
{1-\frac{\nu}{\Lambda}\xi _{{\bf k}_{\nu}+{\bf q}_{eh}}}
\times 
$$
\begin{equation} \label{Pi-theta}
\times  U_{1,|\xi _{{\bf k}_{\nu}+{\bf q}_{eh}}|}
(\theta_1,\theta,\theta_3)
U_{2,|\xi _{{\bf k}_{\nu}+{\bf q}_{eh}}|}(\theta_4,\theta,\theta_2)\; .
\end{equation}
${\bf k}_{\nu}$ is the momentum of a particle at angle $\theta$ with energy
$\xi =\nu \Lambda$.  ${\cal J}(\epsilon,\theta)\equiv
J[(x,y)/(\epsilon,\theta)]$ is the Jacobian of the transformation from
rectangular coordinates in momentum space to the polar ones. 
$X$ is the exchange operator defined as $XU(\theta_1,\theta_2,\theta_3)=
U(\theta_2,\theta_1,\theta_3)$. $U_1$ and $U_2$ in the expression 
(\ref{Pi-theta}) symbolize $U$ or $XU$ as given in ref. \cite{ZS_prb}.
Note that the
interactions $U$, $U_1$, and $U_2$ in the expressions (\ref{Xi-theta}) and
(\ref{Pi-theta}) are not at the actual cutoff $\Lambda$, but at a cutoff 
given by the
configuration of external legs and the integration momentum ${\bf k}_{\nu}$
(entering via total momentum ${\bf q}_{ee}={\bf k}_1+{\bf k}_2$ and momentum
transfer ${ \bf q}_{eh}={\bf k}_1-{\bf k}_3$).  To solve the renormalization
equations we have discretized the $\theta$--dependence of the interaction,
approximating in that way the function $U_{\Lambda}(\theta _1, \theta _2,
\theta _3)$ by a set of coupling constants
$U_{\Lambda}(i_1,i_2,i_3)$. Fig.\ref{Figure 2} shows the flow of some of the
93 coupling constants for the case of a discretization in 16 patches. There
is one pole at $l_c=\ln(8t/\Lambda_c)$, determined by the initial
interaction and the chemical potential, and in this case it is in the
parquet regime.  All coupling constants which have logarithmic corrections
diverge for $l=l_c$.  We identify this point as the critical temperature
$T_c=8t \exp (-l_c)$ in a mean field sense.  The critical temperature at
half--filling, $T_c^0$ is slightly inferior to the mean-field one, but the
ratio $l_c^{MF}/l_c=0.985$ does not depend on the strength of the
interaction. As one sees in fig.\ref{Figure 5}, $T_c$ is finite for any
filling, in contrast to mean--field calculations where $T_c$ falls to zero
at a threshold doping given approximately by $|\mu| =T_c^0$.

To determine the dominant fluctuations at $T=T_c$, we calculate the
correlation matrices for antiferromagnetism $\chi ^{SDW}(\theta _1, \theta
_2)$ and for superconductivity $\chi ^{SC}(\theta _1, \theta _2)$, defined
as
$$
\chi _{\bf q}^{\delta}(\theta_1,\theta_2;|\tau _1-\tau_2|)
=
\int _>d\epsilon _1 \int _>d \epsilon_2 \; {\cal J}(\epsilon_1,\theta_1)
 {\cal J}(\epsilon_2,\theta_2) 
$$
\begin{equation} \label{Correl-funct}
\langle \hat{\Delta}_{\bf q}^{\delta}(\epsilon_1,\theta_1; \tau_1)
\bar{\hat{\Delta}}_{\bf q}^{\delta}(\epsilon_2,\theta_2; \tau_2)
\rangle,
\end{equation}
with $\delta=SC,SDW$. The symbols ``$>$'' mean that the energy integrals are 
over energies {\it outside} of the shell $\pm \Lambda$.
Consequently, $\chi ^{SC}$ and $\chi ^{SDW}$ have the  interpretation of the 
susceptibilities at the temperature $T=\Lambda$. At the beginning of the 
renormalization when $\Lambda=\Lambda_0=8t$ both susceptibilities are zero.

 The order parameter variables are
\begin{equation} \label{SC-def}
\hat{\Delta}_{\bf q}^{SC}(\epsilon,\theta; \tau)\equiv \sum
_{\sigma}\sigma\Psi_{\sigma,{\bf k}}(\tau)
\Psi_{-\sigma,-{\bf k}+{\bf q}}(\tau),
\end{equation}
\begin{equation} \label{SDW-def}
\hat{\Delta}_{\bf q}^{SDW}(\epsilon,\theta; \tau)\equiv \sum
_{\sigma}\bar{\Psi}_{\sigma,{\bf k}}(\tau)
\Psi_{-\sigma,{\bf k}+(\pi,\pi)+{\bf q}}(\tau),
\end{equation}
%\begin{figure}
%\epsfig{file=./Fig3.eps,width=9cm,height=3cm}
%\caption{
%\narrowtext 
%The diagrams for renormalization of SDW correlation function. Note that the 
%propagators in the loop are not at the same energy. For the superconducting 
%channel, the diagrams are similar, except that the loops are p-p, and both 
%lines of the loops are at at the energy $\Lambda$.}
%\label{Figure 3}
%\end{figure}
where {\bf k} is given by the angle $\theta$ and the energy $\epsilon$.  We
consider only the static and long--wavelength limit and follow the
renormalization of the maximal eigenvalues of both correlation matrices; in
the SDW channel the corresponding eigenvector belongs to the $A_1$
representation ($s$--wave), while superconductivity has $B_1$ symmetry
($d_{x^2-y^2}$--wave).  For low doping (small $\mu$) the divergence at
$\Lambda=T_c$ occurs in the magnetic channel (the ``SDW'' region in fig.
\ref{Figure 5}).
 For higher doping, the pairing susceptibility diverges first, i.e. the
``d--wave'' region is a $d_{x^2-y^2}$ superconductor.  The triple point
metal--SDW--superconductor is at the crossover line $T\approx |\mu|$. As we
are considering a two--dimensional system here one should be careful about
the interpretation of $T_c$: in the case of magnetism, this indicates the
onset of well--defined finite--range correlations. For weak interactions,
this is typically a very well--defined crossover.\cite{schulz_89} In the
case of pairing $T_c$ can be identified with the onset of quasi--long--range
order.

In our calculations we have neglected self-energy corrections which are in
Polchinski's formalism given by Hartree--Fock like terms with renormalized
$\omega$ and {\bf q} dependent vertices. The broadening and redistribution
of the spectral weight of the quasiparticles is then determined by the
dynamics of the vertex, which is irrelevant in lowest order by power
counting and is therefore neglected. One should however notice that at the
two--loop order self--energy effects do become important, as known from the
one--dimensional case.\cite{Rev1} In that sense, our $T_c$ should be
understood as a temperature where the effects of interaction start to change
strongly not only the two--particles correlations, but the single particle
properties as well.

Using the above results and arguments, a more detailed description of the
phase diagram, fig.\ref{Figure 5}, can be given: the transition between
Fermi liquid and the magnetic phase occurs in the parquet regime.  Here the
precursor effects, divergence of the commensurate SDW correlation function
and relevance of umklapp processes, suggest that $T_c$ is the temperature of
a metal-insulator transition occurring together or very close to a magnetic
instability. For a more specific description of this phase one should
include self--energy terms. In the pairing phase (d--wave in fig.\ref{Figure
5}) one expects a spin-gap, due to the formation of d--wave singlets, as one
sees from the renormalization the pairing correlations. The maximum pairing
temperature is about $T_c^0/4$, in fact quite close to conserving
approximation calculations \cite{bickers_89} and to the experimental value
of about 1/6 in lanthanum high--$T_c$ compounds.\cite{books} The
$d_{x^2-y^2}$ symmetry of the pairing is also in agreement with the majority
of experiments in cuprates. \cite{d-wave}

From large N arguments\cite{Shankar,ZS_prb} we know that the self--energy
corrections will disappear as $T_c/t$ if $T_c$ is deeply in the BCS
regime. Consequently, far from half filling\cite{ZS_prb}, $T_c$ is a very
good approximation for a superconducting transition, stabilized already by
an infinitesimal inter--plane hopping. Finally, mean-field
arguments\cite{these} suggest that one expects an incommensurate SDW only in
the BCS regime, where perfect nesting is impossible. However, the precision
of our calculation (we cut the Brillouin zone into up to 24
$\theta$-patches) is not sufficient to check whether a magnetic correlation
function diverges at $T_c$ at some incommensurate wave vector.

In conclusion, we have investigated the two--dimensional Hubbard model using
a perturbative renormalization group approach. This allows us to treat
instabilities in the particle--particle and particle--hole channels on an
equal footing and in particular to address the important question of
superconductivity induced by spin fluctuations. In the vicinity of
half--filling we do indeed find a sizeable superconducting transition
temperature, about one fourth of the typical magnetic energy scale. The
pairing is of the $d_{x^2-y^2}$ type.
 
\begin{acknowledgments}
We thank P. Nozi\`eres for important comments. D. Z. thanks J. Schmalian for
interesting discussions and K. H. Bennemann for his hospitality at the
Institut f\"ur Theoretische Physik der Freien Universit\"at Berlin.
\end{acknowledgments}

%\end{multicols}
\newpage


\begin{references}
\bibitem{JerS} D. J\'erome and H. J. Schulz, Adv. Phys., {\bf 31},  299
(1982); D. J\'erome, in ``Organic Conductors'', ed. J. P. Farges (Marcel
Dekker, New York, 1994), p. 405.

\bibitem{Takigawa} M. Takigawa, H. Yasuoka and G. Saito,
J. Phys. Soc. Jpn. {\bf 56}, 873 (1987).

\bibitem{Hadrienetal} J. E. Schirber {\em et al.}, Phys. Rev. B {\bf 44},
4666 (1991); H. Mayaffre {\em et al.}, Europhys. Lett. {\bf 28},
205 (1994); K. Miyagawa {\em et al.}, Phys. Rev. Lett. {\bf 75}, 1174 (1995).

\bibitem{mu-sr}D. R. Harshman {\em et al.}, Phys. Rev. Lett. {\bf 64}, 1293
(1990); J. P. Le {\em et al.}, Phys. Rev. Lett. {\bf 68}, 1923 (1992);
H. Mayaffre {\em et al.}, Phys. Rev. Lett. {\bf 75}, 4122 (1995).

\bibitem{books} ``{\em Physical Properties of High temperature
Superconductors}'', vol. 1--5, ed. by D. M. Ginsberg (World Scientific,
Singapore); ``{\em High Temperature Superconductivity: The Los Alamos
Symposium 1989}'', ed. by K. S. Bedell {\em et al.} (Addison Wesley, redwood
City, 1990).

\bibitem{d-wave} D. A. Wollman {\em et al.}, Phys. Rev. Lett. {\bf 74}, 797
(1995); J. R. Kirtley {\em et al.}, Nature (London) {\bf 373}, 225 (1995);
Z.-X. Shen {\em et al.}, Science {\bf 267}, 343 (1995); R. C. Dynes, Solid
State Commun., {\bf 92}, 53 (1994); W. N. Hardy {\em et al.},
Phys. Rev. Lett. {\bf 70}, 3999 (1993); J. A. Martindale {\em et al.},
Phys. Rev. {\bf B47}, 9155 (1993); D. J. Scalapino, Phys. Rep. {\bf 250},
329 (1995).

\bibitem{Anderson_87} P. W. Anderson, Science {\bf 235}, 1196 (1987).

\bibitem{Shankar} R. Shankar, Rev. Mod. Phys. {\bf 66}, 129 (1994).

\bibitem{ZS_prb} D. Zanchi and H. J. Schulz, Phys. Rev. B {\bf 54}, 9509
(1996).

\bibitem{Rev1}V. J. Emery, in ``{\em Highly Conducting One--Dimensional
Solids}'', eds. J. T. Devreese, R. P. Evrard and V. E. Van Doren (Plenum, New
York 1979), p. 247; J. S\'olyom, Adv. Phys.  {\bf 28}, 201 (1979).

\bibitem{Bourbonnais} C. Bourbonnais and L. G. Caron,
Int. J. Mod. Phys. {\bf 5}, 1033 (1991).

\bibitem{DzYak} I. E. Dzyaloshinskii, Sov. Phys.  JETP {\bf 66}, 848 (1987);
I. E. Dzyaloshinskii and V. M. Yakovenko, Sov. Phys.  JETP {\bf 67}, 844
(1988).

\bibitem{HJS} H. J. Schulz, Europhys. Lett. {\bf 4}, 609 (1987).

\bibitem{Ma} S. K. Ma, Modern Theory of Critical Phenomena, Benjamin (1976).

\bibitem{Polchinski84} J. Polchinski, Nucl. Phys. {\bf B231}, 269 (1984).

\bibitem{Morris94} T. R. Morris, Int. J. Mod. Phys. {\bf A9}, 2411 (1994). 

\bibitem{these} D. Zanchi, Ph.D. thesis, Universit\'e Paris-Sud (1996) 
(unpublished).

\bibitem{schulz_89} H. J. Schulz, Phys. Rev. B {\bf 39}, 2940 (1989).

\bibitem{bickers_89} N. E. Bickers, D. J. Scalapino, and S. R. White,
Phys. Rev. Lett. {\bf 62}, 961 (1989).
\end{references}
\end{document}